\documentclass[prl,twocolumn,a4paper,superscriptaddress]{revtex4}

\usepackage{color}
\usepackage[squaren,Gray]{SIunits}
\usepackage{graphicx}
\usepackage{sidecap}

\usepackage{graphicx}
\usepackage{bm}
\usepackage[T1]{fontenc}
\usepackage{amsmath,amssymb}
\usepackage{times}
\usepackage{ulem}
\usepackage{color}

\begin{document}

\title{Video-rate large-scale imaging with Multi-Z confocal microscopy}

\author{Amaury Badon}
\affiliation{Department of Biomedical Engineering, Boston University, Boston, Massachusetts 02215, USA}
\author{Seth Bensussen}
\affiliation{Department of Biomedical Engineering, Boston University, Boston, Massachusetts 02215, USA}
\author{Howard J. Gritton}
\affiliation{Department of Biomedical Engineering, Boston University, Boston, Massachusetts 02215, USA}
\author{Mehraj R. Awal}
\affiliation{Department of Physiology and Biophysics, Boston University School of Medicine, Boston, Massachusetts 02218, USA}
\author{ Christopher V. Gabel}
\affiliation{Department of Physiology and Biophysics, Boston University School of Medicine, Boston, Massachusetts 02218, USA}
\affiliation{Boston University Photonics Center, Boston, Massachusetts 02215, USA}
\author{Xue Han}
\affiliation{Department of Biomedical Engineering, Boston University, Boston, Massachusetts 02215, USA}
\affiliation{Boston University Photonics Center, Boston, Massachusetts 02215, USA}
\author{Jerome Mertz}
\affiliation{Department of Biomedical Engineering, Boston University, Boston, Massachusetts 02215, USA}
\affiliation{Boston University Photonics Center, Boston, Massachusetts 02215, USA}

\date{\today}

%
%

\begin{abstract}

Fast, volumetric imaging over large scales has been a long-standing goal in biological microscopy. Scanning techniques such as fluorescence confocal microscopy can acquire 2D images at high resolution and high speed, but extending the acquisition to multiple planes at different depths requires an axial scanning mechanism that drastically reduces the acquisition speed. To address this challenge, we report an augmented variant of confocal microscopy where the key innovation consists to use a series of reflecting pinholes axially distributed in the detection plane, each one probing a different depth within the sample. As no axial scanning mechanism is involved, our technique provides simultaneous multiplane imaging over fields of view larger than a millimeter at video-rate. We demonstrate the general applicability of our technique to neuronal imaging of both \textit{Caenorhabditis elegans} and mouse brains \textit{in-vivo}. 

\end{abstract}

\maketitle

\section{Introduction}

Recently, there has been a trend toward the development of microscopy systems that can monitor cellular activity over large spatial scales with high temporal resolution. General conditions for such imaging are that the microscope provide enough spatial resolution to distinguish individual cells, and enough temporal resolution to accurately track the sample dynamics of interest. Moreover, to monitor large populations of cells \textit{in-vivo}, volumetric imaging is required, ideally encompassing hundreds to thousands of individual cells. The confluence of these conditions poses a challenge \cite{ji2016technologies,yang2017vivo}.
\begin{figure*}
\centering
\includegraphics[width=0.9\textwidth]{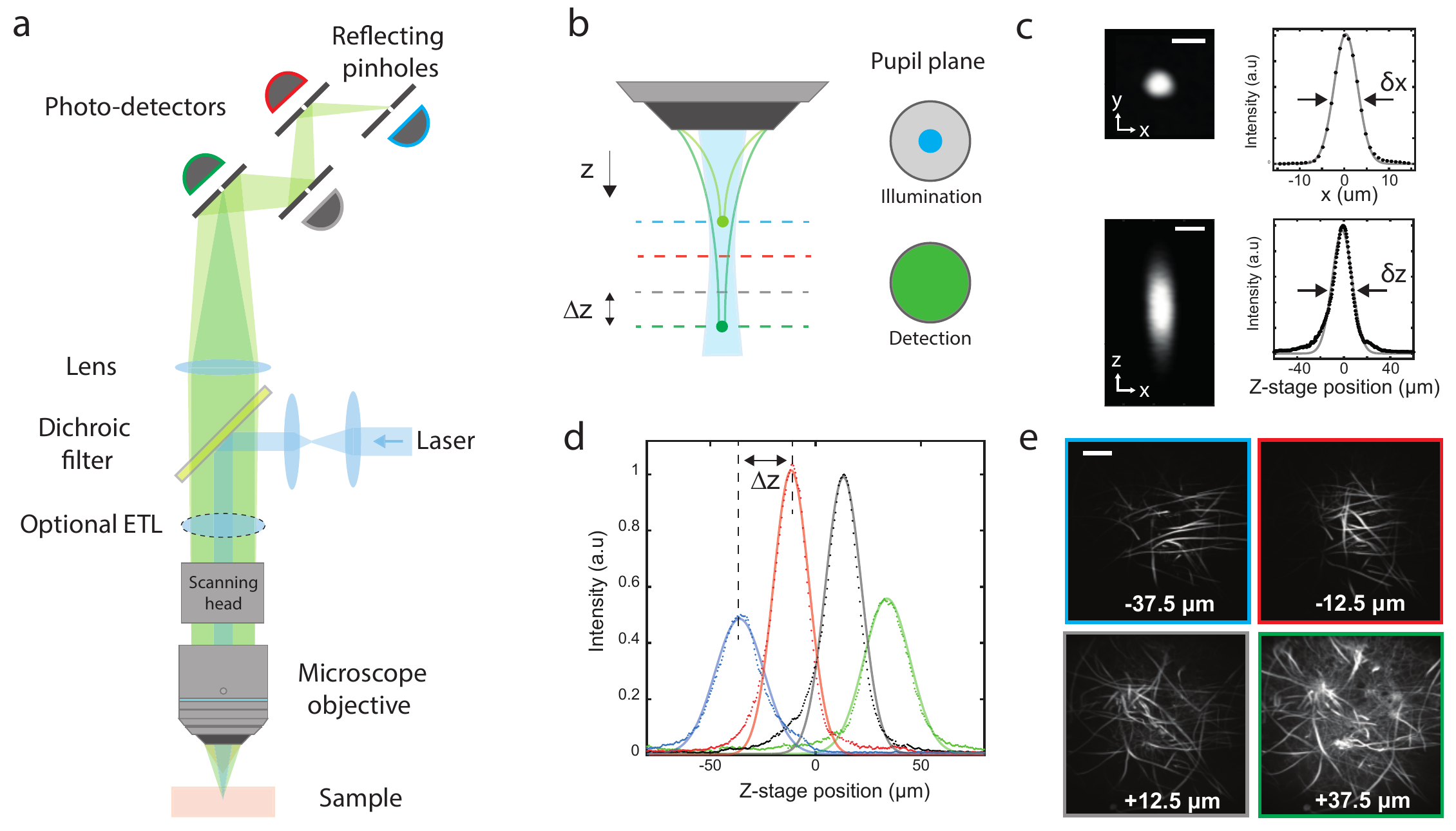}    
\caption{Multi-Z confocal microscopy. (a) Simplified schematic of the experimental setup. The multiplane detection unit is comprised of a series of axially distributed reflecting pinholes, each probing a different depth within the sample. (b) Axially extended illumination is obtained by underfilling the back aperture of the MO. The full NA of the MO is used for detection.  (c) Transverse (xy) and axial (xz) PSF measured with a sub-diffraction size bead, and associated x and z profiles. Scale bar, 5 $\mu$m. (d) Bead signal recorded by each detection channel at different z positions of the stage. Continuous lines correspond to Lorentzian fits. (e) Different imaging planes simultaneously acquired of Aspergillus Conidiophores. Scale bar, 200 $\mu$m.}
\label{fig1}
\end{figure*}
Many previous approaches have been developed for volumetric imaging. For example, widefield microscopes benefit from the large sensor areas and excellent spatiotemporal resolution of modern cameras, and provide single shot 2D imaging. Volumetric imaging at high speed can then be achieved with fast axial scans \cite{botcherby2007aberration,chen2017high,shain2017extended} or instantaneously using wavefront coding \cite{dowski1995extended,quirin2013instantaneous}, diffractive elements \cite{blanchard1999simultaneous,dalgarno2010multiplane,abrahamsson2012fast} or light-field approaches \cite{prevedel2014simultaneous,pegard2016compressive}. However, lack of optical sectioning limits these approaches to shallow depths, sparse samples or imposes a reliance on intensive numerical post-processing. Light-sheet \cite{huisken2004optical,bouchard2015swept} or targeted illumination \cite{xiao2018video} improves sectioning by reducing out-of-focus excitation, but suffers from an inability to prevent scattered background light from impinging the camera sensor, yielding limited image contrast in thick tissue such as mouse brain. 
Alternatively, higher contrast in thick tissue can be obtained with scanning techniques involving multiphoton excitation or confocal detection, but these are generally operated with micron-scale resolution, causing them to be slow and to produce massive amounts of data when operated over large fields of view, or to be limited to modest volumes \cite{kong2015continuous}. Random access techniques \cite{otsu2008optical,nadella2016random} reduce the amounts of data production, but require scan pre-calibration and are sensitive to motion artifacts. Alternatively, data production can be reduced by purposefully limiting the spatial resolution, either axially \cite{yang2016simultaneous,lu2017video} (though see \cite{song2017volumetric,roider20173d,shain2018axial}, where axial resolution can be recovered in sparse samples by post-processing) or near-isotropically \cite{schrodel2013brain}. In our case, since we are primarily interested in the segmentation of individual cells in 3D, near-isotropic resolution is more appropriate.

Here we describe a novel microscopy technique that provides video-rate, multiplane, optically sectioned imaging over large FOVs on the millimeter scale. Our technique, called Multi-Z confocal microscopy, is based on three main ideas. The first is to combine high-NA detection with low-NA illumination. The former leads to high signal collection efficiency; the latter leads to axially extended illumination over a large range of Z depths. The second idea is to detect multiple signals from this extended depth range using multiple confocal pinholes that are axially distributed (similar to the technique in \cite{yang2012z}, though over a much larger scale). The pinholes are reflecting, so that signal rejected by one pinhole is sent to the next pinhole, and so forth. In this manner, no signal is lost, and signal collection efficiency remains high (a reflecting pinhole has been used previously, though in the context of two-photon microscopy \cite{hu2018simultaneous}). The third idea is to exploit the benefits obtained from our larger confocal probe volumes, namely larger signals and avoidance of oversampling at cellular resolution, to scan over much larger FOVs than provided by standard confocal microscopes, thus optimizing our Multi-Z microscope for fast, large-scale imaging of cell populations.  

\section{Multi-Z Principle}

A schematic of our experimental set up is shown in Figure \ref{fig1}a. In practice, our instrument resembles a standard confocal microscope, though with differences in both the illumination and detection optics. To achieve low-NA illumination, we significantly underfill the back aperture of our microscope objective (MO) by inserting an afocal beam compressor in the excitation laser path, which results in a highly axially elongated illumination beam in the sample. With NA$_{ill} \approx$ 0.1, the axial extent on the illumination focus is of the order of 100 $\mu$m. The fluorescence from the sample, in turn, is epi-collected through the full aperture of the same MO, which ensures both maximized fluorescence collection efficiency and a detection depth of field much shorter than the illumination axial depth (see Fig. \ref{fig1}b). The fluorescence is then routed to not one but several pinholes (here four), each conjugate to different depths separated by $\Delta z$ (here 25 $\mu$m) within the illumination profile. The size of the pinholes is adjusted to be approximately $M\delta x$, where $\delta x$ is the transverse illumination width and $M$ is the magnification from the sample to the pinholes (here $M=62.5$). In the image space, the separation between the pinholes $\Delta Z$ is given by approximately $M^2\Delta z$. Because of its quadratic dependence on $M$, this separation is in the range of several centimeters even though $\Delta z$ is in the range of microns, making the pinhole layout particularly convenient to build (as opposed to the more restricted implementation involving high-NA illumination and close-packed micro-machined reflectors \cite{yang2012z}). We note that downstream channels are subject to a small amount of screening from the pinholes in the upstream channels. An evaluation of this minor effect is discussed in \href{link}{Supplement 4}. 

\begin{figure*}[htbp]
\centering
\includegraphics[width=0.85\textwidth]{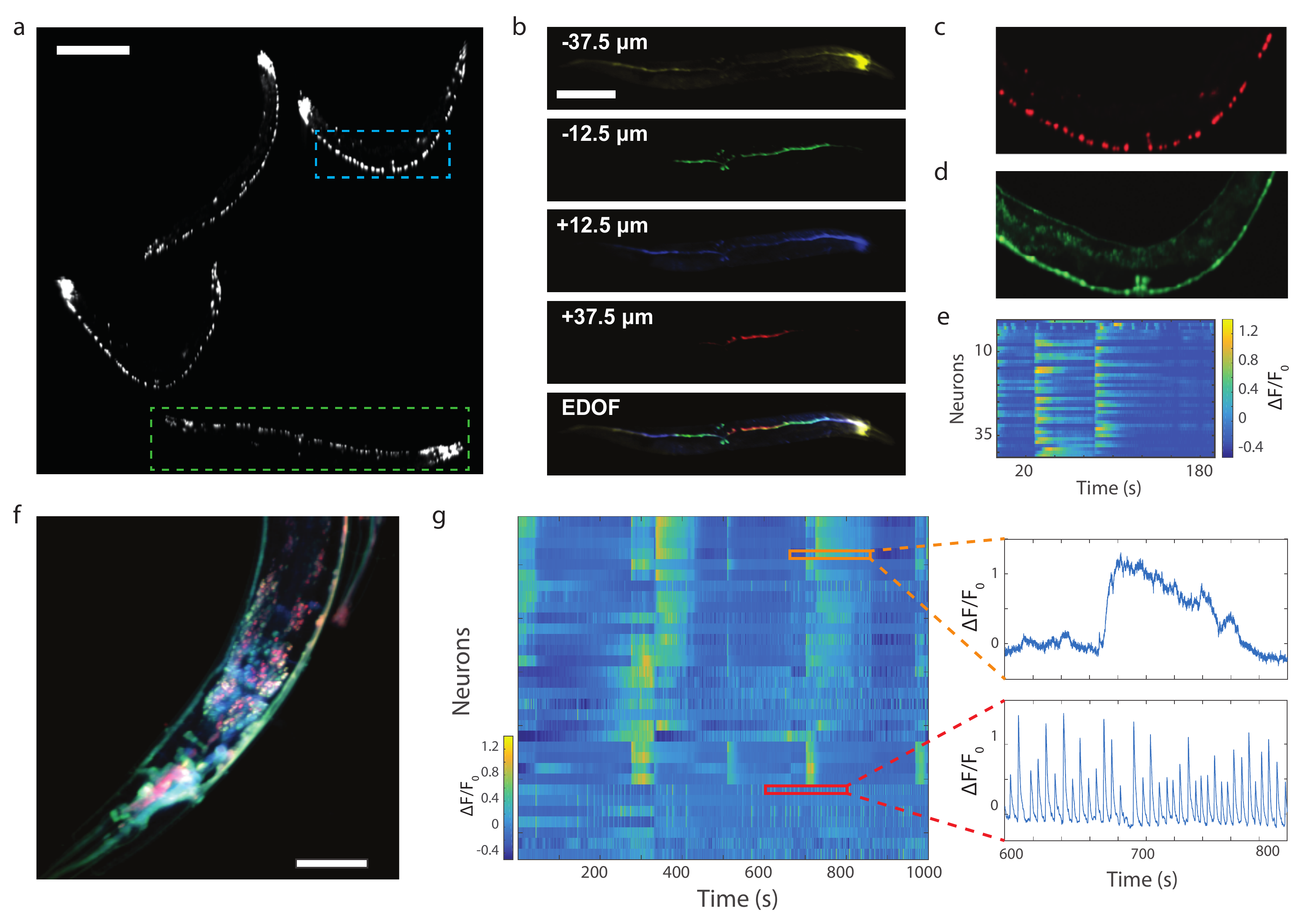}
\caption{In vivo volumetric imaging of \textit{C. elegans}. (a)  Simultaneous imaging of a single worm at different depths and the resulting volumetric rendering. Scale bar, 200 $\mu$m. (b) Extended depth of field image showing the nuclei marked with NLSmCherry of four different worms in the FOV. Scale bar, 200 $\mu$m. (c,d) Close up view (blue rectangle), showing the nuclei (in red) and the neurons displaying activity (in green). (e) Activity of the 42 neurons identified in (d). (f) Standard deviation of the calcium activity recorded in the head ganglion of a worm in each imaging plane. Each channel is displayed with a different color and in log scale. (g) Activity of the 32 neurons identified in (f) and recorded over 1000 s. Each row corresponds to a time-series of an individual neuron, as shown in the insets.  }
\label{fig2}
\end{figure*}

To evaluate the resolution of our microscope, we imaged sub-diffraction-sized beads over a 200 $\mu$m axial range about the MO focal plane. Using a $10\times$ objective, we achieved for each detection channel a resolution of $\delta x$= 4.8 $\mu$m and $\delta z$=15 $\mu$m in the transverse and axial dimensions, respectively (Fig. \ref{fig1}c). Note that this compromise in resolution was purposefully chosen to facilitate high-speed imaging over large FOVs. Such 3D resolution is easily adequate to resolve individual cells, such as neurons in brain tissue, while avoiding problems associated with oversampling and the unnecessary processing of massive amounts of data \cite{weisenburger2017quantitative}. The resultant bead signal simultaneously recorded by the four detection channels clearly demonstrates the multiplane capability of our approach (see Fig.\ref{fig1}d). As expected, the four curves are similar and axially shifted by $\Delta z$=25 $\mu$m, thus illustrating confocal detection over 100 $\mu$m. The slightly broader and lower amplitude responses associated with the first and last detection channels arise from the axial roll-off of our Gaussian-Lorentzian illumination beam away from its nominal focus, as detailed in \href{link}{Supplement 3}. 

 As an example demonstration, we imaged a fixed sample of Aspergillus Conidiophores, a common mold. Figure \ref{fig1}e shows the four simultaneously acquired images from the shallowest ($z=-37.5~\mu$m) to the deepest plane ($z=+37.5~\mu$m). Different features and the branching of the hydrae are clearly visible in these images, demonstrating the simultaneous multiplane acquisition and optical sectioning capabilities of our microscope over a large FOV, here 1.2 $\times$ 1.2 $\text{mm}^2$.

\section{High-speed volumetric imaging}

Because Multi-Z confocal microscopy provides volumetric imaging without the need for axial scanning, its speed is limited by the mechanism for transverse 2D scanning only. In our case, we use standard resonant-nonresonant galvanometric scanning (see Methods) to achieve video rate imaging, easily fast enough to monitor the dynamics of cell activity reporters such as the calcium sensor GCaMP6. In particular, we demonstrate the speed capacity of our system by  monitoring large-scale \textit{in vivo} neuronal activity in different organisms.

\subsection{Whole animal study : \textit{Caenorhabditis elegans}}

We first performed imaging of \textit{C. elegans} expressing both NLSmCherry, targeted to nuclei, and GCaMP6s. Adult worms are typically 500-800 $\mu$m long and 100 $\mu$m wide which makes them difficult to image entirely with a conventional confocal microscope. As a result, only young specimens or the head ganglia are usually imaged. However our microscope readily captured full 3D volumes of multiple worms embedded in agarose gel, here up to four. Using two-color illumination and interleaved acquisition (see Methods) we were able to simultaneously detect the neurons and monitor their activity. Figure \ref{fig2}a displays the extended depth of field (EDOF) image of the nuclei obtained by summing the intensity of the four channels. Numerous neurons of the worms, from the densely packed head ganglia to the tail are visible. Since the calcium activity is simultaneously recorded, the active neurons can be spatiotemporally resolved in 4D (see Fig.\ref{fig2}b). Such a rendering would not be possible with conventional confocal microscopy, which is limited to single-plane imaging over typically much smaller FOVs. Moreover, our approach enabled us to perform volumetric imaging of moving specimens (see Visualization 1). Because of the agarose gel, the worm was constrained to rotations only but our approach could be extended to freely moving worms as well. From the same acquisition, the $\text{Ca}^{2+}$ activity of the 42 detected ventral neurons of a selected worm (blue dashed rectangle) was monitored at 7.5 Hz for 200 s (Fig. \ref{fig2}.d). The corresponding $\text{Ca}^{2+}$ traces are shown in Fig. \ref{fig2}e, where different subsets of neurons reveal clearly correlated and anti-correlated activity associated to forward and backward motion, in agreement with previous observations \cite{schrodel2013brain,prevedel2014simultaneous}. 
\begin{figure*}[htpb]
\centering
\includegraphics[width=1\textwidth]{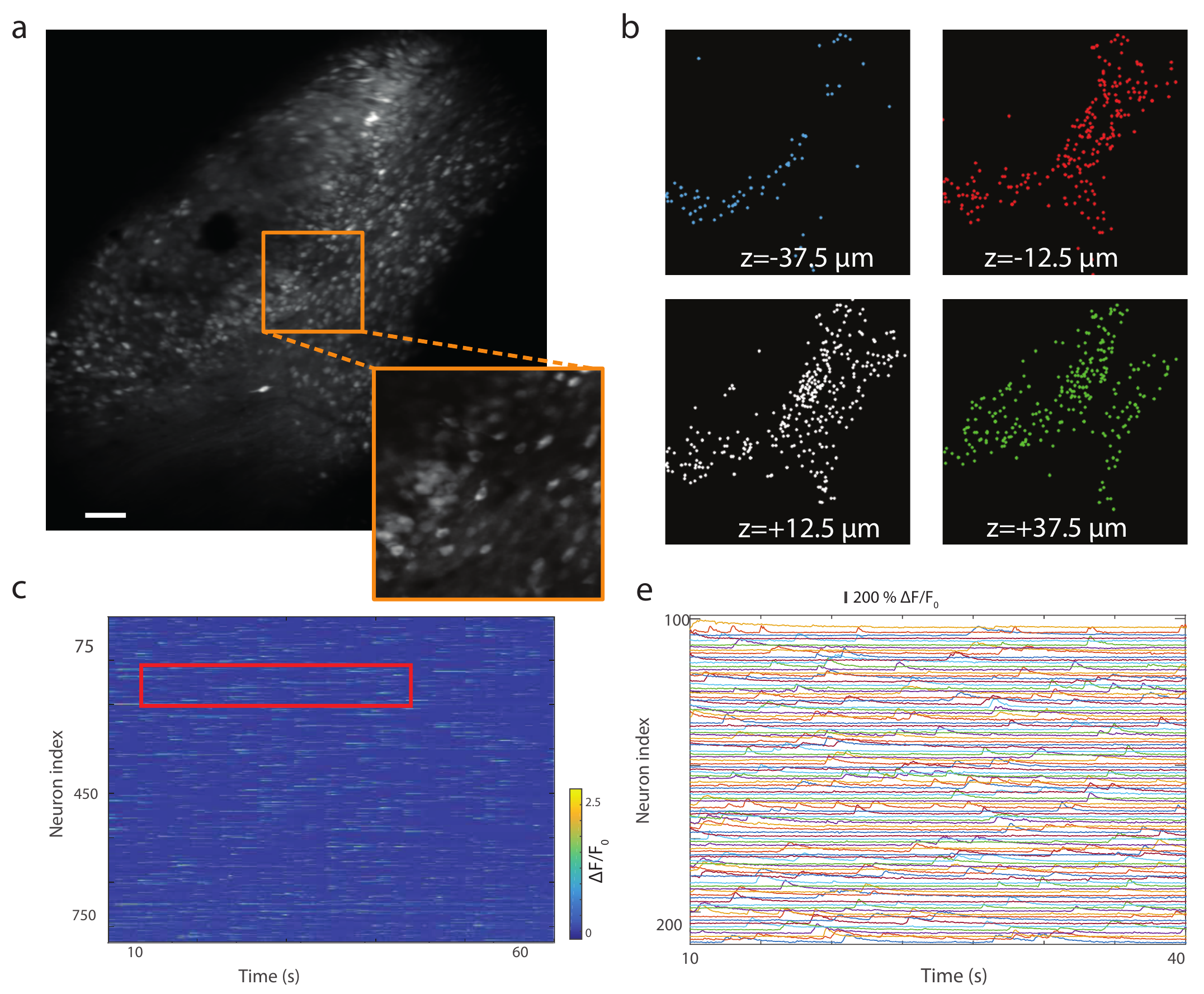}
\caption{Video-rate volumetric $\text{Ca}^{2+}$ imaging in mouse brain expressing GCaMP6f (note that viral injection here led to a labelled area somewhat smaller than our FOV) . (a) Extended depth of field image recorded in the hippocampus at 30 frames per second and averaged over 20 seconds. Scale bar, 100 $\mu$m. Inset with $3\times$ zoom illustrates cellular resolution. (b) Identification of neurons in each image plane using constrained non-negative matrix factorization. (c) Activity of the 826 neurons identified in (a). (d) Magnified view of the neuronal activity traces for the region indicated by the red rectangle in (c).}
\label{fig3}
\end{figure*}

To image densely packed regions such as head ganglia, a higher transverse resolution is preferable. This could be obtained while maintaining the same distance $\Delta$z between imaging planes by simply switching the MO (20$\times$) and two relay lenses to maintain a constant $M$ (see Methods). In this new configuration, we imaged the head ganglion of a worm (zoomed to 300 $\times$ 300 $\times$ 100 $\mu \text{m}^3$) at 30 Hz. Figure \ref{fig2}f shows an overlay of the temporal standard deviation of the $\text{Ca}^{2+}$ signals, where color corresponds to depth. Neurons from different depths are clearly distinguishable in this densely populated region. Note that due to internal motion, the worm intestine is also highlighted when we compute the standard deviation. In total, 32 neurons are identified over a duration of 1000 s, and their $\text{Ca}^{2+}$ traces are presented as a heat map in Fig. \ref{fig2}g. Our acquisition speed was amply sufficient to record even the fastest $\text{Ca}^{2+}$ dynamics (see also Visualization 2).

\subsection{Mouse brain}

More challenging is the demonstration of \textit{in vivo} neuronal imaging in mammalian brains. For this, we performed $\text{Ca}^{2+}$ imaging in a mouse brain expressing GCaMP6f, requiring both fast imaging, at least 15 Hz, and optical sectioning to reject extraneously scattered light. We focused in particular on the hippocampus, a subcortical region of the brain (see Methods). During imaging, the mice were awake, head-fixed and no sensory stimulus was applied. 

We recorded the spontaneous $\text{Ca}^{2+}$ activity of neurons within a $1200 \times 1200 \times 100$ $\mu \text{m}^3$ volume at 30 Hz, the fastest rate currently achievable with our system for this FOV. Though the labelling was confined to an area smaller than our FOV, the EDOF images (i.e. z projections) reveal a large population of neurons (see figure \ref{fig3}a). Higher resolution images of smaller regions of interest were also obtained by reducing the FOV while keeping the same acquisition speed and number of pixels, allowing the details of individual cells to become clearer (see inset). The benefits of simultaneous, multiplane acquisition are apparent in Fig. \ref{fig3}b, where different neurons are revealed in different imaging planes. Using a constrained non-negative matrix factorization algorithm (CNMF) \cite{pnevmatikakis2016simultaneous}, 90, 264, 312 and 260 neurons were separately identified in each plane, from deepest to shallowest respectively, resulting in a total number of 926 neurons. Owing to the intentional partial overlap of the image planes (see Fig. \ref{fig1}d), some of these identifications are expected to be redundant. Indeed, similar processing applied to the EDOF images pared back these identifications to a total of 826 independent neurons, indicating that the overlap-induced crosstalk was approximately 12\% (Fig. \ref{fig3}a and Visualization 3). The corresponding extracted $\text{Ca}^{2+}$ traces, recorded over 66 s, are shown in Figs. \ref{fig3}c and \ref{fig3}d for a magnified view. These results illustrate the capacity of our system to provide adequate SNR for neuronal segmentation in large, relatively dense, populations at video-rate timescales.

\section{Augmented volumetric imaging}
\begin{figure*}[htbp]
\centering
\includegraphics[width=1\textwidth]{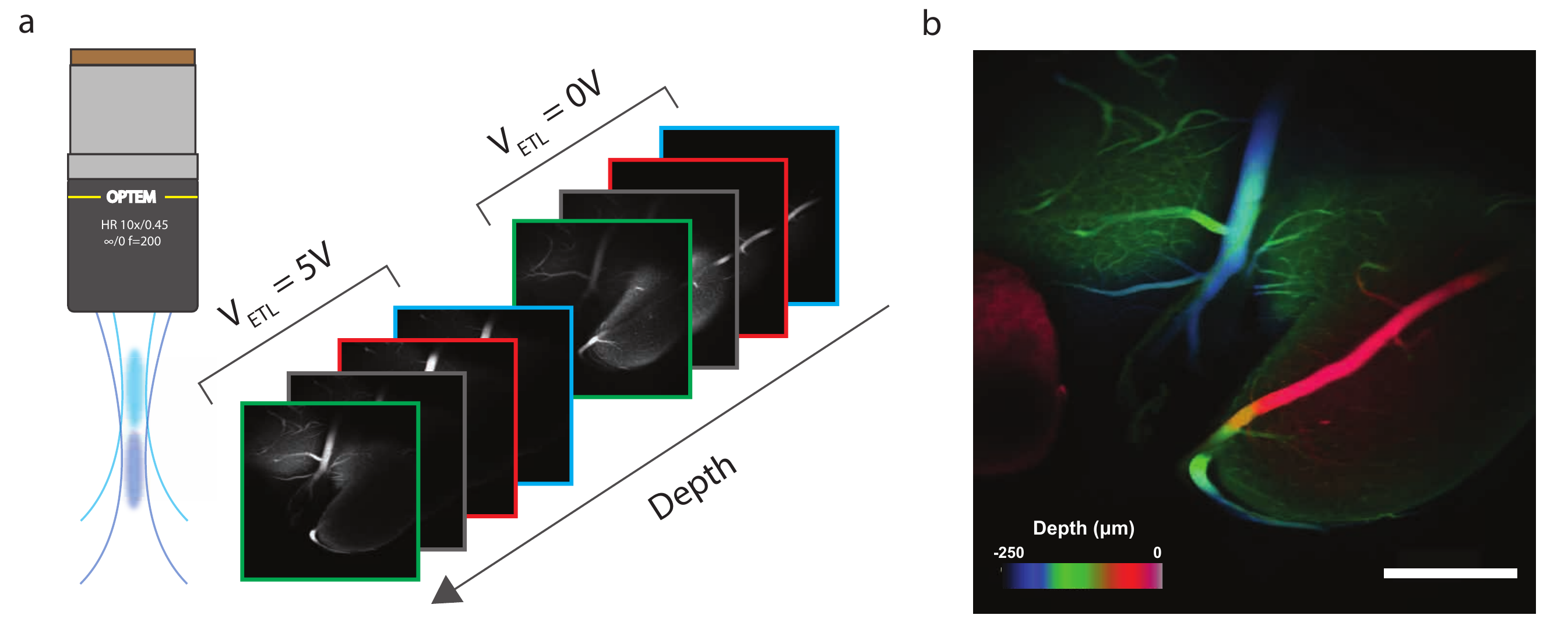}
\caption{Augmented volumetric imaging using an electrically tunable lens controlled by a square-wave voltage $V_{ETL}$. (a) Multi-Z images obtained for $V_{ETL}=0$ and $V_{ETL}=5V$ corresponding to a focal shift of 150 $\mu$m. (b) Resulting volumetric acquisition of fixed mouse brain vasculature with structures color-coded by depth. Scale bar 300 $\mu$m.}
\label{fig4}
\end{figure*}

Our microscope was designed to provide multiplane acquisition over an axial range of about 100 $\mu$m. In the event that a larger axial range is desired, several strategies can be considered. The first, and simplest, is to further extend the range of the illumination focus by further underfilling the back-aperture of the MO. In this case, the detection channels can be spread more sparsely along the illumination profile, reducing the detection fill factor and hence the spatial overlap of neighboring detection probe volumes. Alternatively, to maintain the same fill factor, the number of channels (and pinholes) can be increased. Such a simple solution, however, would incur a loss in transverse spatial resolution. For example, a doubling of the axial range would lead to an worsening of the transverse resolution by a factor $\sqrt{2}$.   

Yet another strategy that does not impair transverse resolution involves the use of a mechanism to rapidly change focal depths. Example mechanisms are an electrically tunable lens (ETL) \cite{grewe2011fast}, an acoustic gradient lens \cite{olivier2009two,duocastella2014simultaneous} or a deformable mirror \cite{shain2017extended}. In our case, we demonstrated the augmentation of our axial range with an ETL (Optotune VIS-EL-10-30-C), which we inserted in the vicinity of the pupil plane between the scanning mirrors and the MO. By controlling the focal strength of the ETL, we thus axially translated both the illumination and detection foci by up to 150 $\mu$m, while incurring little change in the magnification and resolution of our system. In particular, a square-wave control voltage was applied to the ETL at half the microscope frame rate, leading to interleaved stack acquisition that doubles the effective axial range of our microscope, though at the cost of halving our net acquisition speed. 

As an example demonstration, we imaged a fixed mouse brain with FITC-labeled vasculature. In a two-step acquisition process, we recorded eight frames in total, four from each axial position of the ETL, leading to an overall imaging volume of $1200 \times 1200 \times 250~\mu \text{m}^3$ acquired at a net speed of 15 Hz. Figure \ref{fig4}b displays a resulting volume projection with color-coded depth, revealing both larger vessels that span the full axial extent of the recorded volume and smaller vessels visible only in individual frames. These preliminary results further demonstrate versatility of our microscope in providing large-scale imaging.

\section{Summary}

In summary, we have developed a generalized version of confocal microscopy that provides simultaneous multiplane imaging over large FOVs while remaining fast, here video rate. Advantages of our system are that it is highly light efficient, for two reasons. First, it makes full use of the detection NA and incurs no loss upon signal detection through reflecting pinholes, thus maximizing signal collection efficiency. Second, whereas in a standard confocal microscopy the illumination light produces only a single image plane at a time, in Multi-Z confocal microscopy the \textit{same} illumination light is repeatedly utilized to produce multiple image planes at a time, thus making more efficient usage of the laser excitation power and, concomitantly, for the acquisition of equal image stacks, reducing the deleterious effects associated with this excitation power, such as photobleaching, photodamage, etc.. This second excitation efficiency advantage is expected to scale with the number of pinholes. 

Other advantages of Multi-Z microscopy are that the system requires no scan pre-calibration and no image post-processing. Indeed, the images shown here were not post-processed in any way, save for image registration to correct for sample dither, illustrating the robustness of our system against motion artifacts which can easily be monitored in all three dimensions. On the other hand, post-processing could be envisaged, such as 3D deconvolution, as facilitated by the direct acquisition of volumetric image data. Also, our approach is robust against axial motion artifacts that may arise, for example, from blood flow or animal breathing, which is a common problem in \textit{in vivo} applications. While in conventional confocal microscopy axial motion can lead to total loss of the signal, here the signal simply appears in a different channel.

Still other advantages are that our system is readily scalable to a greater number of imaging planes, with no speed penalty in principle, by simply adding pinholes and detectors, or to an augmentation of axial range with no penalty in transverse resolution. Moreover, it is versatile in that it is amenable to multi-color imaging (e.g. two-color here), and $M$ and/or $\Delta z$ can be adjusted with only minor optical modifications. For example, since $\Delta Z= M^2 \times \Delta z$, one can adjust the distance between the imaging planes $\Delta z$ by simply tuning the total magnification of the system $M$, or alternatively by tuning the physical distance between the pinholes $\Delta Z$. In our case, $\Delta z$ was adjusted so that the confocal detection volumes filled the axial extent of the illumination beam and there was no gap between image planes. Other geometries are, or course, possible.  

The simplicity and ease of use of our system should make it attractive for general biomedical research applications.

\section{Methods}

\subsection{Hardware setup}

Light from a blue (488nm/80mW, Omicron PhoxX) or yellow-green laser (561nm/50mW, Vortran Stradus) was scanned by a pair of galvanometric mirrors (CRS 8 kHz, Cambridge Technology) and relayed by lenses to fill only a fraction of the back pupil aperture of the MO (Optem LWD 10X, NA=0.45 or Olympus XLUMPLFLN-W 20X NA=0.95). The underfilling of the pupil results in low NA illumination (NA$_{ill}\approx$ 0.1) that provides an axially extended focus in the focal plane of the MO. The fluorescence signal was epi-detected by the full NA of the same MO, ensuring maximal collection efficiency and constrained axial resolution. The fluorescence signal was then de-scanned and isolated with a dichromatic mirror (403/497/574, Edmund Optics) and two notch filters (OD4 488NM, Edmund Optics and NF03-561E-25, Semrock). After being relayed and further magnified, the fluorescence was detected by four reflecting pinholes (d=300 $\mu$m, National Aperture) that were axially distributed. The separation between consecutive pinholes was 10 cm, which, given the magnification of the system (M = 62.5$\times$), corresponds to a separation of $\Delta$z = 25 $\mu$m in the object space. The signals from the four photo-detectors (MicroFC-SMA-30020, SensL) positioned behind the pinholes were amplified by custom built voltage amplifiers and digitized by a high-speed 4-channel FPGA board (PXI-7961R, National Instruments). The setup was controlled from a dual-CPU workstation and the experiments were conducted using Scanimage \cite{pologruto2003scanimage} microscope control software. 

\subsection{Fixed samples experiments}

PSF measurements were performed using 0.2 $\mu$m fluorescent beads (TetraSpeck, ThermoFisher T14792) mounted on a glass slide. We used a commercially available Aspergillus Conidiophores glass slide (Carolina Biological Supply Co. 297872). An average laser output power of 3mW was used for these experiments.

\subsection{\textit{Caenorhabditis elegans} preparation}

\textit{C. elegans} strains were cultivated at 20$^\circ$C following standard procedures (on NGM agar seeded with Escherichia coli OP50 as a food source). All imaging experiments were performed on young-adult hermaphrodites using the transgenic strain QW1217 (zfIs124[Prgef-1::GCaMP6s];otIs355[Prab-3::NLS::tagRFP]) expressing panneuronal GCaMP and nuclear-localized RFP in all neurons in the worm. Worms were partially immobilized for imaging by following a hydrogel encapsulation protocol \cite{burnett2018rapid}. In brief, worms were placed into a droplet of solution consisting of 10 PEG-DA and 0.1$\%$ Igracure. The droplet containing the worms was then hardened by a minute-long exposure to UV-light, which crosslinked the solution into a gel, preventing gross movement of the worms. An averaged laser output power of 5mW  was used for these experiments.

\subsection{Mouse preparation and Imaging}

All animal procedures were in accordance with the National Institutes of Health Guide for the care and use of laboratory animals and approved by the Boston University Institutional Animal Care and Use Committee. Female C57BL/6 mice (n=2, Taconic; Hudson, NY), 8-12 weeks old at the time of surgery, were first injected with  AAV9-Syn-GCaMP6f.WPRE.SV40 virus obtained from the University of Pennsylvania Vector Core (titer $\sim$6e12 GC/ml). 250 nL of virus was stereotaxically injected into the CA1 region (AP: –2mm, ML: 1.4 mm, DV: –1.6 mm) using a 10 nL syringe (World Precision Instruments) fitted with a 33 gauge needle (NF33BL; World Precision Instruments), at a speed of 40 $\mu$L/min controlled by a micro-syringe pump (UltraMicroPump3–4; World Precision Instruments). Upon complete recovery, animals were surgically implanted with custom imaging windows, which consisted of a stainless steel cannula (OD: 0.317 cm., ID: 0.236 cm., height 2 mm) adhered to a circular coverslip (size 0; OD:3mm) with a UV-curable optical adhesive (Norland Products). After careful aspiration of the overlying cortical tissue, using the corpus callosum as an anatomical guide, the imaging window was placed above the CA1 viral injection site. During the same surgery, a custom aluminum head-plate was attached to the skull anterior to the imaging cannula for head fixation during imaging. Animals were imaged beginning 4 weeks later to allow for full expression of GCamp6f. Prior to imaging for the first time, animals were also habituated to a custom mount head fixation apparatus over several sessions as previously described \cite{mohammed2016integrative}. An average laser output power of 5 mW was used for these experiments. Imaging sessions lasted typically 15-20 minutes.

\subsection{Fixed-brain preparation}

Before extracting a mouse brain, a cardiac perfusion with heparinized saline followed by FITC-albumin-gelatin was performed. After extraction, the brain was immersed in a 4\% PFA solution for 6 hours and then in a PBS solution for 3 days. Subsequently, the solution was replaced by distilled water with 0.5\% alpha-thioglycerol and increasing amounts of fructose, 20, 40, 60, 80 and finally 100\%.

\subsection{Image processing and data analysis }

Motion correction was performed using the Moco plugin \cite{dubbs2016moco} in ImageJ. Data analysis was carried out using custom scripts in Matlab (MathWorks) for the \textit{C. elegans} experiments and using constrained non-negative matrix factorization algorithm for the mice experiments \cite{pnevmatikakis2016simultaneous}. To extract $\Delta F$/$F_0$, we computed $\Delta$ F/$F_0$  (t) = 100 × (F(t)-$F_0$)/$F_0$, $F_0$ being the mean fluorescence intensity of each trace. 

\section*{Funding Information} 
This work was partially supported by NIH grant R21-EY027549 and the Wallace H. Coulter Foundation.

\section*{Acknowledgments} 
We thank Timothy Weber, Jean-Marc Tsang and Sheng Xiao for technical help and helpful discussions. We thank the BU Neurophotonics Center for general support, and especially Kivilcim Kilic for providing the fixed brain sample. 

\bibliography{optica}

\end{document}